\documentclass[aps,prl,twocolumn,preprintnumbers,10pt,showpacs,nofootinbib]{revtex4-1}
\usepackage{amsmath,amsfonts,epsfig,color,latexsym}
\usepackage{amssymb}
\usepackage{epsfig,psfrag}

 \usepackage{slashed}
\usepackage[normalem]{ulem}
\usepackage{color}

\begin{document}
\renewcommand{\thefootnote}{\fnsymbol{footnote}}
\newpage
\pagestyle{empty}
\setcounter{page}{0}


\newcommand{\norm}[1]{{\protect\normalsize{#1}}}
\newcommand{\p}[1]{(\ref{#1})}
\newcommand{\half}{\tfrac{1}{2}}
\newcommand \vev [1] {\langle{#1}\rangle}
\newcommand \ket [1] {|{#1}\rangle}
\newcommand \bra [1] {\langle {#1}|}
\newcommand \pd [1] {\frac{\pa}{\pa {#1}}}
\newcommand \ppd [2] {\frac{\pa^2}{\pa {#1} \pa{#2}}}
\newcommand{\ed}[1]{{\color{red} {#1}}}

\newcommand{\cI}{{\cal I}}
\newcommand{\cM}{{\cal M}} 
\newcommand{\cR}{{\cal R}} 
\newcommand{\cS}{{\cal S}} 
\newcommand{\cK}{{\cal K}}
\newcommand{\cL}{{\cal L}} 
\newcommand{\cF}{{\cal F}}
\newcommand{\cN}{{\cal N}}
\newcommand{\cA}{{\cal A}}
\newcommand{\cB}{{\cal B}}
\newcommand{\cG}{{\cal G}}
\newcommand{\cO}{{\cal O}}
\newcommand{\cY}{{\cal Y}}
\newcommand{\cX}{{\cal X}}
\newcommand{\cT}{{\cal T}}
\newcommand{\cW}{{\cal W}}
\newcommand{\cP}{{\cal P}}
\newcommand{\bP}{{\bar\Phi}}
\newcommand{\mK}{{\mathbb K}}
\newcommand{\nt}{\notag\\} 
\newcommand{\pa}{\partial}
\newcommand{\ep}{\epsilon}
\newcommand{\om}{\omega}
\newcommand{\bom}{\bar\omega}
\newcommand{\etap}{\bar\epsilon}
\newcommand{\vep}{\varepsilon}
\renewcommand{\a}{\alpha}
\renewcommand{\b}{\beta}
\newcommand{\g}{\gamma}
\newcommand{\s}{\sigma}
\newcommand{\la}{\lambda}
\newcommand{\tl}{\tilde\lambda}
\newcommand{\tm}{\tilde\mu}
\newcommand{\tk}{\tilde k}
\newcommand{\da}{{\dot\alpha}}
\newcommand{\db}{{\dot\beta}}
\newcommand{\dg}{{\dot\gamma}}
\newcommand{\dd}{{\dot\delta}}
\newcommand{\q}{\theta}
\newcommand{\bq}{{\bar\theta}}
\renewcommand{\r}{\rho}
\newcommand{\br}{\bar\rho}
\newcommand{\be}{\bar\eta}
\newcommand{\bQ}{\bar Q}
\newcommand{\bx}{\bar \xi}
\newcommand{\tx}{\tilde{x}}
\newcommand{\tr}{\mbox{tr}}
\newcommand{\+}{{\dt+}}
\renewcommand{\-}{{\dt-}}
\newcommand{\ti}{{\textup{i}}}


\preprint{CERN-TH-2018-069, LAPTH-012/18, MPP-2018-42, MITP/18-24}

\title{Amplitudes from superconformal Ward identities}

\author{D.\ Chicherin$^a$, J.\ M.\ Henn$^{a,b}$, E.\ Sokatchev$^{a,c,d}$}

\affiliation{
$^a$ 
PRISMA Cluster of Excellence, Johannes Gutenberg University, 55099 Mainz, Germany\\
$^b$ Max-Planck-Institut f{\"u}r Physik, Werner-Heisenberg-Institut, 80805 M{\"u}nchen, Germany\\
$^c$ LAPTh, Universit\'e Savoie Mont Blanc, CNRS, B.P. 110, F-74941 Annecy-le-Vieux, France\\
$^d$ Theoretical Physics Department, CERN, 1211 Geneva 23, Switzerland}


\begin{abstract}
We consider finite superamplitudes of ${\mathcal N=1}$ matter, and use superconformal symmetry to derive powerful first-order differential equations for them. 
Due to on-shell {collinear singularities}, the Ward identities have an anomaly, which is obtained from lower-loop information. 
We show that in the five-particle case, the solution to the equations is uniquely fixed by {the expected analytic behavior}. 
We apply the method to a non-planar two-loop five-particle integral.
\end{abstract}

\maketitle

Conformal symmetry has played a central role in quantum field theory for many decades.
In recent years, its consequences are being actively explored within the AdS/CFT correspondence, the bootstrap program, and high-energy QCD.

In particle physics, scattering amplitudes are fundamental objects, relevant for
collider physics. In the high-energy regime the masses can often be neglected,
and the Lagrangian becomes conformal. However, putting the external particles on shell can render the symmetry anomalous.
Tree-level amplitudes have a holomorphic anomaly \cite{Witten:2003nn,Cachazo:2004by,Cachazo:2004dr,Bidder:2004tx} 
that arises when external particles become collinear.
This mechanism is responsible for the breakdown of (super)conformal symmetry of tree-level amplitudes and of discontinuities 
of loop amplitudes \cite{Bargheer:2009qu,Korchemsky:2009hm,Beisert:2010gn}.

The conformal symmetry of finite loop integrals and the associated anomalous Ward identities were studied in ref. \cite{Chicherin:2017bxc}.
The anomaly occurs when an external light-like momentum becomes collinear with a loop momentum.
The equations provide non-trivial constraints; however, due to their second-order nature, it is in general not straightforward to solve them.

In the present paper, we show how to obtain powerful first-order differential equations in a model of $\cN=1$ massless supersymmetric matter. 
The amplitudes are infrared finite. A  certain sector of the amplitudes consists of ultraviolet finite Feynman diagrams, so within it we are not affected by the running coupling. 
Naively, the Feynman integrals in this sector should enjoy all the symmetries of the model, in particular (super)conformal symmetry. 
We show that special conformal supersymmetry is broken by collinear contact terms inside the finite Feynman integrals, and we derive Ward identities that quantify this effect. We remark that a similar phenomenon was observed in the context of the supersymmetric Wilson loop in refs. \cite{CaronHuot:2011ky,Bullimore:2011kg,CaronHuot:2011kk}.

As a first application, we focus on the five-particle case, which is interesting for several reasons. First, it turns out that for four particles (super)conformal symmetry does not give any restrictions. So, from this point of view, five particles is the first non-trivial case. On the other hand, five-particle scattering at higher loops involves intricate transcendental functions \cite{Gehrmann:2015bfy,Papadopoulos:2015jft,Chicherin:2017dob} and is of considerable current interest \cite{Badger:2017jhb,Abreu:2017hqn}. 

We show how the differential equations can be used to fully determine the answer. 
The fermionic generator is reduced to the so-called twistor collinearity operator of \cite{Witten:2003nn}.
Its kernel consists of a function of holomorphic cross-ratios only \cite{Korchemsky:2009hm}.
We propose a way to fix this boundary freedom by imposing expected analytic properties of the function. 
We illustrate these ideas by computing a non-trivial non-planar two-loop five-particle integral.

\section{$\cN=1$ matter superamplitudes and their 
superconformal anomalies } 

We consider scattering amplitudes in the Wess-Zumino model of $\cN=1$ massless supersymmetric matter.
The multiplet is described by a chiral superfield $\Phi(p,\q)
$ and its antichiral conjugate $\bar\Phi(p,\bq)$. 
The Lagrangian contains cubic  interactions $g\int d^2\q \Phi^3$ and $g \int d^2\bq\bar\Phi^3$. 

\begin{figure}
\begin{center}
\includegraphics[width = 6cm]{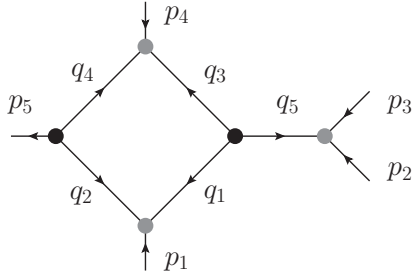}
\end{center}
\caption{One-loop $\mathcal{N} = 1$ matter supergraph. Black and grey blobs denote antichiral and chiral vertices, respectively.}\label{figuresupergraphs1}
\end{figure}

\begin{figure}[t]
\begin{center}
\includegraphics[width = 6cm]{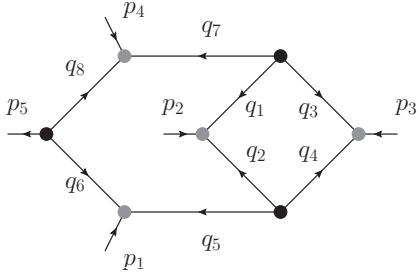}
\end{center}
\caption{Non-planar two-loop $\mathcal{N} = 1$ matter supergraph.} \label{figuresupergraphs2}
\end{figure}

This model is superconformal at the classical level. At the quantum level, the symmetry is broken, but only
by propagator corrections, and the beta function is proportional to the anomalous dimension of the superfield \cite{Ferrara:1974fv}.
This property allows us to study individual finite supergraphs, with the only requirement that they do not contain propagator correction subgraphs. See Figs.~\ref{figuresupergraphs1} and \ref{figuresupergraphs2} for sample graphs. 

Such graphs are naively superconformal. We will see that the symmetry is in fact broken by on-shell collinear effects. 
The latter can be controlled and give rise to powerful anomalous Ward identities. 

In order to discuss superamplitudes, we introduce the on-shell super-states
\begin{align}\label{1}
& \bar\Phi(p,\eta) = \bar\phi(p) + \eta \psi_-(p)\, , \;\; \Psi(p,\eta) =  \psi_+(p) + \eta \phi(p)  
\end{align}
They depend on the light-like momentum $p_{\a\da} = \la_\a \tl_\da$, and on the Grassmann variable $\eta$. For the antichiral on-shell state $\eta =[\tl\bq] \equiv \tl_\da \bq^\da$; for the chiral state it is defined as the Fourier transform of $\la^\a \q_\a$.  
The Feynman rules of the quantum theory are well known. 
Here we only need the mixed propagator (wave function) 
$\vev{\bar\Phi(-p,\bq) \Psi(p,\eta)} = \eta+ [\tl\bq]$, and the antichiral cubic vertex involving a Grassmann integral, $g \int d^2\bq\, \bar\Phi^3$. 


In our $\cN=1$ model, the 
breaking of conformal supersymmetry
can already be seen for the three-point vertex function of two off-shell superfields and one on-shell state. Apart from illustrating the mechanism, this object will also constitute the main building block for our practical calculations.
At leading order, it is given by
\begin{align}
\cF  \equiv & \vev{\bar\Phi(q_1,\bq_1) \bar\Phi(q_2,\bq_2)|\bar\Phi(p,\eta)}_g \nonumber \\
=&  \delta^4(P)\,  \delta^2(Q) \, \frac{g}{q^2_1 q^2_2}  \,.  \label{4}
\end{align}
The delta functions account for momentum $P=q_1+q_2+p$ and supercharge $Q=\bq_1 q_1 + \bq_2 q_2 +\eta\la$ conservation. Invariance under $\bar Q = \sum_i \frac\pa{\pa \bq_{i}} +  \tl   \frac\pa{\pa\eta}$ is also readily verified.  
We focus on the chiral superconformal generator $ S_\a = \frac{1}{2} \sum_i \frac{\pa^2}{\pa q_i^{\a\da}\pa\bq_{i\,\da}}  + \frac{\pa^2}{\pa\eta \pa \la^\a}  $. It annihilates 
$\cF$ for generic momentum configurations. However, this is not true in the collinear regime $q_1\sim q_2 \sim p$. When the bosonic derivatives in $S_\a$ act on the product of propagators in \p{4}, they generate contact terms. We postpone the details to a future publication and give the result for the {\it collinear superconformal anomaly} 
\begin{align}\label{2.19}
S^\a  \cF  =& 2i\pi^2 \la^{\a}\, \int_0^1 d\xi\, \left(\eta + [\tl\bar\q_1]\xi + [\tl\bar\q_2] \bar\xi \right) \, \nonumber \\
& \qquad\qquad \times \delta^{4}(q_1+\xi p)\, \delta^{4}(q_2+\bx p) \,,
\end{align}
where $\bx=1-\xi$. The antichiral generator  is not anomalous, $\bar S_\da \cF=0$. The superconformal algebra $\{S_\a, \bar S_\da\}= K_{\a\da}$ then yields an anomaly of the conformal generator $K$ similar to that of \cite{Chicherin:2017bxc}.

The anomalous Ward identity \p{2.19} tells us that when acting on an on-shell supergraph, such as the one shown in Fig.~\ref{figuresupergraphs1}, we pick up an anomaly contribution from each chiral three-point vertex (grey blob) connected to an external antichiral on-shell state. Thanks to the extra delta function in (\ref{2.19}), a loop integration is localized, so that the anomaly term for an $L$-loop graph is expressed in terms of a one-fold integral of an $(L-1)$-loop graph.

Let us now apply this anomaly equation to computing $\cN=1$ matter superamplitudes.
They can be classified according to their Grassmann degree.
Due to $Q$ supersymmetry, the general form of an amplitude of $m$ $\bar\Phi$ states and $n$ $\Psi$ states is
\begin{align}
{\cal A}_{n,m} = \delta^{4}(P) \delta^2(Q) \mathcal{P}_{n,m}\,.
\end{align}
The Grassmann degree of $\mathcal{P}_{n,m}$ is related to the $U(1)$ R-symmetry of the theory.
We have $R_\eta = 1$, $R_{\bar\Phi} = 2/3$, and $R_\Psi = 1/3$. Therefore, $\mathcal{P}_{n,m}$  has $R$-charge $(n+2m)/3-2 \ge 0$. 
In this letter, we consider the five-particle case $n=1$ and $m=4$, i.e. $R_{\mathcal{P}}=1$, as a first application.

The five-leg amplitudes we consider have the structure
\begin{align}\label{5leg}
{\cal A}_{1,4} =& \vev{\bar\Phi(p_1,\eta_1)  \bP(p_2,\eta_2) \bar\Phi(p_3,\eta_3)\bar\Phi(p_4,\eta_4) \Psi(p_5,\eta_5) } \nonumber \\
=& \delta^{4}(P) \, \delta^2(Q) \, \Xi_{123} \,  C(p) \,.
\end{align}
Here we used the $\bar Q$ invariant
\begin{align} \label{XIijk}
\Xi_{123} = \eta_1 [23] + \eta_2 [31] + \eta_3 [12]\,.
\end{align} 
Any three $\eta$'s define such an invariant, but all the choices are equivalent due to supercharge conservation, e.g., $\Xi_{123} = \frac{\vev{45}}{\vev{12}} \Xi_{345}$, etc. 
Thus, the entire superamplitude \p{5leg} is determined by a {\it single bosonic function} $C(p)$. 
The latter can be found by extracting a component amplitude, for example, 
$[23] \vev{45} C =  \vev{\psi_{-}(1) \bar{\phi}(2) \bar{\phi}(3)  \psi_{-}(4) \phi(5)} $, which corresponds to the $\eta_1 \eta_4 \eta_5$ term in \p{5leg}.

Let us now turn to  the superconformal Ward identities. 
When $S^\a$ acts on the RHS of eq.~(\ref{5leg}), it commutes with $ \delta^{4}(P) \, \delta^2(Q)$, so that we only need the relation
\begin{align}\label{58}
S^\a [\Xi_{123} C ] = F^{\a}_{123}C\,,
\end{align}
where 
\begin{align}\label{XiEquiv}
F^\a_{123} = [23]\frac{\pa}{\pa \la_{1\, \a}} +[31]\frac{\pa}{\pa \la_{2\, \a}} + [12]\frac{\pa}{\pa \la_{3\, \a}} 
\end{align}
is the so-called {\it twistor collinearity operator}, see \cite{Witten:2003nn}. 
We remark that we could have equally well acted with $\bar{S}^{\dot \alpha}$ on the conjugated amplitude ${\cal A}_{4,1}$, which has $R_{\mathcal{P}}=0$.

In order to determine the right-hand side of the Ward identity we need to evaluate explicitly the $S$-variation of the supergraphs contributing to ${\cal A}_{1,4}$ (recall that we neglect propagator corrections). 
The chiral vertex functions are exactly $S$-invariant. 
The antichiral ones are invariant up to a contact term, see \p{2.19}, 
which becomes relevant due to the loop integrations. 
So for ${\cal A}_{1,4}$ we have contributions from the  antichiral legs 1,2,3,4.
The generator $S^\a$ lowers the Grassmann degree by one, so we have
\begin{align}\label{SA}
S^\a  {\cal A}_{1,4} = \delta^{4}(P) \, \delta^2(Q)\,  \sum_{i=1,2,3,4} \la^\a_{i} A_{i}(p) \,  
\end{align}
with some bosonic {anomaly functions} $A_{i}(p)$. Comparing with eqs.~(\ref{5leg}) and \p{58}, we derive  the {\it anomalous superconformal Ward identity}
\begin{align}\label{SWI}
F_{123}^\a C(p) = \sum_{i=1,2,3,4} \la^\a_{i} A_{i}(p)\,.
\end{align}
{We note that the freedom in choosing the superinvariant $\Xi_{ijk}$  also affects the collinearity operator $F_{ijk}$. Thus we have
$\vev{45} F_{345} C = \vev{12} F_{123} C$, etc., where momentum conservation is assumed. However, the additional equations do not provide new information.}

{It is convenient to define the dimensionless, helicity-neutral function 
\begin{align}\label{11}
  f = \langle 45 \rangle [14][23]\, C \,.    
\end{align}
For general five-particle kinematics, $f$} depends on four dimensionless variables.
They can be chosen as
\begin{align}\label{defx1x2}
x_1= -1-\frac{s_{14}}{s_{15}}\,, \qquad  x_2= -1-\frac{s_{14}}{s_{45}}
\end{align}
where $s_{ij}=2 p_i \cdot p_j$, and
\begin{align}\label{defx3x4}
x_3 = \frac{ [12][34]}{[23][41]} \,,\quad x_4 = \frac{[23][45]}{[34][52]} \,.
\end{align}
The real variables $x_1,x_2$ are parity even, while the complex variables $x_3,x_4$ undergo conjugation under parity. 

Let us comment on the solutions of the homogeneous equation $F_{123}^\a \tilde{f} =0$. 
The two components of this equation fix the dependence on $x_1, x_2$, while  $F_{ijk}^\alpha x_3 = F_{ijk}^\alpha x_4 = 0$, so
that any function of the holomorphic variables $x_3, x_4$ solves the homogeneous equation, see also \cite{Korchemsky:2009hm}. 
However, transcendental functions of these variables have unphysical branch cut properties, 
consequently we expect the freedom of the homogeneous solution to reduce to just one integration constant.

\section{Box with off-shell leg}

Let us illustrate the method with a one-loop example. We consider the box with one off-shell leg, see Fig.~\ref{figuresupergraphs1}. 
Extracting the component $\eta_1 \eta_4 \eta_5$ we express the bosonic function  $R$ as the following Feynman integral
\begin{align}\label{14}
f = - [14] \int\frac{d^4 \ell}{i \pi^2}\, \frac{\vev{1|q_2 \tilde q_4 |4}}{q_1^2 q_2^2 q_3^2 q_4^2}\,.
\end{align}
As is well known \cite{Dixon:2011ng}, this coincides with the `magic' pentagon integral of \cite{ArkaniHamed:2010gh}, as well as with the six-dimensional one-mass-easy box. It is given by
\begin{align}\label{422}
f = {\rm Li}_2(x_1 x_2 ) - {\rm Li}_2(-x_1 ) - {\rm Li}_2(-x_2 )+  {\rm Li}_2(1 ) \,.
\end{align}

Let us now derive the result for $f$ from the superconformal Ward identity (\ref{SWI}).
Anomalies originate from the antichiral   
legs 1 and 4, see eq. (\ref{2.19}), while $A_2 = A_3 = 0$, up to contact terms.
A short supergraph calculation gives
\begin{align}\label{16}
A_1 = \frac{\log(s_{23}/s_{45}) }{\vev{15}(s_{23}-s_{45})}  \,, \quad
A_4 = \frac{\log(s_{23}/s_{15}) }{\vev{45}(s_{23}-s_{15})}    \,.
\end{align}
Projecting the two components of eq. (\ref{SWI}) with independent spinors,  
we find 
\begin{align}\label{17}
&x_1 \partial_{x_1} f(x_1,x_2)=   \log \frac{1+x_1}{1-x_1 x_2} \equiv a_1(x_1,x_2)  \,,\nt
&x_2 \partial_{x_2} f(x_1,x_2) =   \log \frac{1+x_2}{1-x_1 x_2}\equiv a_4(x_1,x_2)\,.
\end{align} 
It is clear from Fig.~\ref{figuresupergraphs1} that the functions $f, a_1$, and $a_4$ are independent of $x_3, x_4$. 

The system of 
differential equations
\p{17} 
determines $f$ up to an arbitrary constant. The latter can be fixed by demanding finiteness of $C$ in \p{11} as $s_{14} \to 0$.
The final solution takes the form
\begin{align}\label{22}
f  = \int_{-1}^{x_1} \frac{dt}{t}\, a_1(t,x_2)   = \int_{-1}^{x_2} \frac{dt}{t}\, a_4(x_1,t)  \,.
\end{align}
To see that each of these integrals solves both equations in \p{17}, one uses the integrability condition $x_2 \partial_{x_2} a_1  - x_1 \partial_{x_1} a_4  =0$ and the property $a_1(x_1,-1)=a_4(-1,x_2)=0$ of the anomaly terms.
It is easy to check that \p{22} agrees with (\ref{422}).

We find it amusing to note that, although the superconformal Ward identities are trivial for four particles, 
one can nonetheless obtain the result for the four-particle amplitude by taking the (finite) collinear limit $p_{2} \sim p_{3}$ 
of our result, which yields $\log^2 s_{15}/s_{45} + \pi^2$.

\section{Application to non-planar five-particle integral}

Next, we consider the non-planar pentabox diagram of Fig.~\ref{figuresupergraphs2}.
Its  $\eta_1 \eta_4 \eta_5$ component is given by
\begin{align}\label{topC}
f = \frac{-[14][23]}{ (i \pi^2)^2} \int 
\frac{d^4 \ell_1 d^4 \ell_2}{q_1^2 \ldots {q}^2_{8}} \bra{2} q_{2} \tilde{q}_{4}\ket{3} \bra{1} q_{5} \tilde{q}_{7}\ket{4}\,.
\end{align}
The symbol \cite{Goncharov:2010jf} of this integral was recently obtained using the bootstrap approach \cite{Chicherin:2017dob}.
Here we compute the full function, in a straightforward way, using our new method.
In this way, we also prove the result of \cite{Chicherin:2017dob}, without  any assumptions about the symbol alphabet.

The anomalous superconformal Ward identity for this integral involves four anomaly terms $A_{i}$ corresponding to the four chiral vertices in Fig.~\ref{figuresupergraphs2}.
Due to the $1\leftrightarrows 4$ and $2 \leftrightarrows 3$ symmetries, only two of them are independent.
The delta functions in eq. (\ref{2.19}) reduce the $A_{i}$ to one-loop integrals.
$A_1$ is determined by the one-mass box with `magic' numerator computed above; $A_2$ involves a sum of two pentagons with one massive corner.

\begin{widetext}
Using known expressions for these one-loop functions, we find 
\begin{align}\label{eqwidetext}
&A_{1} = \frac{1}{\vev{15}}\int^1_0 \frac{d\xi}{\xi s_{45} + \bar\xi s_{23}}
\left[
{\rm Li}_2\left(1-\frac{\xi s_{45} + \bar\xi s_{23}}{\xi s_{12}}\right) +
{\rm Li}_2\left(1-\frac{\xi s_{45} + \bar\xi s_{23}}{\xi s_{13}}\right) + 
\frac{1}{2} \log^2\left(\frac{s_{12}}{s_{13}}\right) + \frac{\pi^2}{6}
\right] \nt
&A_{2} = - [25] \int^1_0 \frac{d\xi}{ \vev{51}[13]\vev{32}[25] +   \bar\xi s_{23} s_{25}} \log \frac{\xi \bar\xi s_{12} s_{24}}{(\xi s_{15} + \bar\xi s_{34})(\xi s_{13} + \bar\xi s_{45})}  
\log \frac{\bar\xi s_{45}(\xi s_{15} + \bar\xi s_{34})}{\xi s_{15}(\xi s_{13} + \bar\xi s_{45})}\,.
\end{align}
\end{widetext}
Integrating over $\xi$, we find that the anomaly equation \p{SWI} takes the form   
\begin{align}\label{WIforSymb}
F^\a_{123} {f} = [14][23] \vev{45} \sum_{i=1}^4 \la_i^\a \, \frac{a_i}{r_i}\,,
\end{align}
where  $a_i$ are pure functions of weight three, and where
\begin{align} \label{rfun}
r_1 =& {\vev{15}(s_{23}-s_{45})} \,, \qquad  r_2 = {\vev{25}s_{23}} \;,\;\nonumber \\
 r_3 =& {\vev{35} s_{23}} \,, \qquad  
r_4 = {\vev{45}(s_{23}-s_{15})} \,.
\end{align}
Projecting eq. (\ref{WIforSymb}) onto its independent components, we can write
\begin{align}\label{R2looppartial2}
\tilde d {f} =& \, a_1 \, \tilde d \log {x_1} + a_4 \,  \tilde d \log  x_2  \, \\ 
& \hspace{-0.5cm} + a_2  \, \tilde d  \log \frac{1-x_1 x_2 }{ (1+x_2)(x_3-1)x_4+(1+x_1)(x_3 x_4-1) }  \, \nonumber \\
&  \hspace{-0.5cm} + a_3 \, \tilde d \log  \frac{1-x_1 x_2 }{ (1+x_2) x_3 x_4+(1+x_1)(x_3 x_4-1)  }  \nonumber \,, 
\end{align}
with $\tilde d = dx_1 \,\partial_{x_1} + d x_2 \,\partial_{x_2}$. 
{We note that the integrability condition ${\tilde{d}}^2 f = 0$ yields a non-trivial cross-check on the calculation of the anomaly terms $a_{i}$ appearing in eq. (\ref{R2looppartial2}).}
Eq.~(\ref{R2looppartial2}) determines $f$ up to an arbitrary function $g(x_3, x_4)$. We now give two ways of fixing the latter. 

One method follows the one-loop example above. 
Using a Feynman parameter representation of the integral \p{topC}, one can show that $\lim_{s_{14} \to 0}f= f(x_1=x_2=-1)=0$, while keeping $x_3,x_4$ constant, i.e. $g(x_3,x_4)=0$.

This remarkable fact is not a coincidence and leads us to a second method of fixing the boundary data.
Imagine that we have already found the correct solution for $f$. 
We argue that any additional, non-trivial function $g(x_3,x_4)$ would introduce unphysical analytic behavior, such as
branch cuts depending on the holomorphic variables $x_3, x_4$ only.
This argument also explains the existence of some values of $x_1, x_2$ (in our case $(-1,-1)$), for which $f$ and hence $g$ are independent of $x_3, x_4$. 

We can write the solution for $f$ by integrating, e.g., $\partial_{x_1} f$ from the boundary point $x_1 = -1$, in close analogy with eq.~\p{22}, but this time with contributions from  $a_1,a_2,a_3$. To see this, one notices that the parameter representations (\ref{eqwidetext}) satisfy $a_{1}(x_1,-1,x_3,x_4)= a_{4}(-1,x_2,x_3,x_4)=0$.
For convenience, we focus on the kinematic region $s_{i,i+1}<0,s_{13}<0, s_{24}<0$, for which $f$ is real-valued. 
Some care is required, as the boundary point $x_1=-1$ lies outside the above region.
The integration is performed analytically using \cite{Panzer:2014caa} and evaluates to multiple polylogarithms \cite{2001math3059G}.
The final expression is given as an ancillary file.

Let us discuss checks of our result.
The symbol of the result agrees with ref. \cite{Chicherin:2017dob}.
The function has discontinuities \cite{2001math3059G} only at expected values of the $s_{ij}$, and not at values depending on the holomorphic variables $x_3 , x_4$.
Finally, we compared the numeric evaluation \cite{Vollinga:2004sn} to that of an integral representation.



Let us study the symbol of our answer in the light of the structure suggested by the differential equation (\ref{R2looppartial2}), which constrains the possible last entries.
Four of the last entries simply correspond to the arguments of the logarithms of eq. (\ref{R2looppartial2}). 
In terms of the $\{W_{i} \}$-alphabet, $i=1\ldots 31$ of ref.~\cite{Chicherin:2017dob}, they are given by $W_{14}/W_{5}$, $W_{2}/W_{20}$, $W_{2}/W_{18}$ and $W_{12}/W_{4}$. 
Other last entries must be functions of $x_3, x_4 $ only. There are only $5$ such letters in the alphabet of \cite{Chicherin:2017dob}, namely $W_{5}W_{17}W_{26}/(W_{1} W_{4})$, and cyclic. We find that the symbol of our answer can indeed be written in terms of eight of those nine last entries.

\section{Discussion}

We derived powerful superconformal Ward identities for finite scattering amplitudes in a model of ${\mathcal N}=1$ supersymmetric matter. 
The essential reason for the presence of the anomaly lies in the on-shell external legs, which makes singular collinear configurations possible. 


We obtained a first-order differential equation for a given $L$-loop integral, with the right-hand side expressed as a single parameter integral over certain $(L-1)$-loop integrals.
This is to be contrasted with the traditional differential equations approach (see e.g. \cite{Argeri:2007up}), which involves a large system of equations, and whose generation typically requires considerable computer algebra.
Being first-order, our equations are very powerful. Their kernel is easily seen to contain only holomorphic functions.
We argued that the absence of unphysical analytic behavior is enough to fix this freedom.

It is important to note that our method applies equally to planar and non-planar amplitudes. We illustrated this by evaluating a non-planar two-loop five-particle Feynman integral.

The class of integrals considered here is generated by ${\mathcal N}=1$ matter supergraphs without propagator corrections. It is interesting to note that these are closely related to the `local' integrals introduced in \cite{ArkaniHamed:2010gh}. An avenue for future research is to 
extend the present method to include ${\mathcal N}=1$ gauge superfields, which would allow the application to larger classes of integrals. {This will also allow us to study the relationship with the Wilson loop approach of \cite{CaronHuot:2011kk,CaronHuot:2011ky,Bullimore:2011kg}.}

\section{Acknowledgment}

{We are grateful to Simon Caron-Huot for stimulating discussions.} The authors were supported in part by the PRISMA Cluster
of Excellence at Mainz university. This project has received funding from the European
Research Council (ERC) under the European Union's Horizon 2020 research and innovation
programme (grant agreement No 725110), ``Novel structures in scattering amplitudes''.

\bibliographystyle{apsrev4-1} 

\bibliography{bibfile_superconformal}

\end{document}